\documentclass[conference]{IEEEtran}
\usepackage{cite}
\usepackage{amsmath,amssymb,amsfonts}
\usepackage{algorithmic}
\usepackage{graphicx}
\usepackage{textcomp}
\usepackage{multicol}
\usepackage{xcolor}
\usepackage{etoolbox}
\usepackage{caption}
\usepackage{hyperref}
\usepackage{makecell}
\usepackage{float}
\usepackage[pass]{geometry}
\renewcommand{\baselinestretch}{1.00335}\normalsize

\def\BibTeX{{\rm B\kern-.05em{\sc i\kern-.025em b}\kern-.08em
    T\kern-.1667em\lower.7ex\hbox{E}\kern-.125emX}}

\begin{document}

\title{Forecasting Technological Directions in Wireless Networks and Mobile Computing via AutoML Framework \\

\thanks{Identify applicable funding agency here. If none, delete this.}
}

\author{
    \IEEEauthorblockN{
        Ahmed Abolfadl\textsuperscript{1}, 
        Marwa Mahmoud\textsuperscript{2}, 
        Basma Afifi\textsuperscript{1}, 
        Mervat Abu-Elkheir\textsuperscript{1}, 
        Maggie Mashaly\textsuperscript{2}
    }
    \\
    \IEEEauthorblockA{
        \textsuperscript{1}Faculty of Media Engineering and Technology, German University in Cairo, Cairo, Egypt \\
        \textsuperscript{2}Faculty of Information Engineering and Technology, German University in Cairo, Cairo, Egypt
    }
    \\
    \IEEEauthorblockA{
        ahmed.abuelfadel@student.guc.edu.eg, \\
        \{marwa.abla, basma.mohammed, mervat.abuelkheir, maggie.ezzat\}@guc.edu.eg
    }
}

\maketitle

\begin{abstract}
The exponential increase in scientific publications has driven the emergence of new trends. Accurate forecasting of these developments is essential for researchers and professionals to stay updated with advancements in the field. This study presents an automated pipeline for trend prediction in the wireless networks and mobile computing domain by integrating clustering, topic modeling, and time series analysis. The process begins with the collection of 127,820 abstracts from high-impact journals and conferences, followed by extensive preprocessing and semantic embedding using the SPECTER model. AutoCluster applies meta-learning to select the most suitable clustering algorithm based on the dataset meta-features, ensuring semantically coherent groupings. AutoTopicModeling then employs a successive halving strategy to identify the best-performing topic model per cluster, followed by LLM-assisted topic labeling and optional label generalization. Finally, AutoTrendAnalysis transforms topic-labeled data into time series and applies forecasting models -ARIMA, STL, Prophet, or LSTM - to predict future topic popularity. Topics are classified as strong, weak, or noise signals based on forecast trajectories, offering interpretable insights into emerging and declining research themes. The framework is scalable, adaptive, and designed for robust trend analysis across scientific domains. Experimental results demonstrated high predictive accuracy, achieving a Root Mean Square Error (RMSE) of 36.76.
\end{abstract}


\section{Introduction}

The continuous advancement of wireless communications and mobile computing has brought major changes across industries, from smart cities and autonomous systems to next-generation communication services \cite{intro}. As this field evolves at an extraordinary pace, driven by breakthroughs in 5G/6G, mobile edge computing, and IoT, the ability to anticipate future research and technology trends becomes increasingly valuable. Accurately identifying such trends is essential not only for academic researchers and industry innovators but also for policy architects aiming to make forward-looking decisions on infrastructure, security, and regulatory frameworks.

Despite the importance of forecasting developments in this domain, traditional trend analysis methods are often insufficient. The huge volume and heterogeneity of technical publications make manual reviews both time-consuming and prone to bias. Furthermore, the emergence of niche subfields—such as federated learning for mobile systems, UAV-assisted networks, or ultra-low-latency protocols—requires dynamic tools that can adapt to rapid changes in scientific focus and terminology.

To overcome these limitations, this paper proposes an automated machine learning (AutoML)-driven approach to predict trends within wireless networks and mobile computing literature. The proposed framework systematically processes temporal textual datasets to uncover latent thematic patterns and project their trajectory over time, all with minimal manual configuration.

The pipeline is composed of three modular stages. First, AutoCluster determines the most appropriate clustering algorithm for grouping semantically similar documents. Second, AutoTopicModeling selects and tunes topic modeling techniques. Lastly, AutoTrendAnalysis applies the most suitable forecasting model to analyze how each topic evolves over time.
This fully automated system provides a robust, scalable, and interpretable solution for tracking the evolution of ideas and technologies in wireless and mobile research. It enables real-time insights that can inform strategic planning, guide investment in emerging technologies, and foster innovation in a field defined by rapid and constant change.

\section{Literature Review}

Recent advancements in AutoML have focused on automating various stages of the machine learning pipeline, with notable contributions enhancing model selection, hyperparameter tuning, clustering, and trend prediction.

A foundational contribution by He et al. \cite{he2019automl} introduced a comprehensive AutoML framework that automates both model selection and hyperparameter optimization (HPO), achieving superior prediction accuracy and efficiency compared to traditional approaches. Complementing this, Domhan et al. \cite{domhan2015benchmarking} demonstrated the effectiveness of automated HPO in building well-calibrated models, accelerating the development of robust machine learning solutions.

Optuna, introduced by Akiba et al. \cite{akiba2019optuna}, plays a pivotal role in many AutoML systems. It provides a flexible define-by-run interface and incorporates powerful optimization techniques such as Tree-structured Parzen Estimator (TPE) and Covariance Matrix Adaptation Evolution Strategy (CMA-ES). Due to its efficiency and extensibility, Optuna serves as the HPO engine in our system.

In the domain of unsupervised learning, clustering automation has gained traction. The AutoML4Clust framework by Zhao et al. \cite{zhao2022automl4clust} targets efficiency and scalability, leveraging internal validation metrics and efficient search strategies to optimize the full clustering pipeline—particularly in large-scale, high-dimensional scenarios lacking ground truth labels. A broader overview is provided by Huang et al. \cite{huang2023survey}, who survey AutoML techniques for clustering, classifying them by which pipeline stages they automate (e.g., preprocessing, algorithm selection, validation), and highlighting key research challenges such as inconsistent evaluation metrics and diverse data requirements.

Meta-learning has emerged as a key enabler of clustering automation. Bengio et al. \cite{bengio2022evaluating} evaluated a variety of meta-features to identify those most predictive of clustering algorithm performance, emphasizing the critical role of data characterization in effective meta-learning. Building on this, Liu et al. \cite{liu2021clustering} proposed a meta-learning framework for recommending clustering algorithms. Their system uses dataset meta-features to train a meta-learner capable of predicting algorithm performance, thus facilitating data-driven selection without manual tuning. Similarly, Lacoste et al. \cite{lacoste2019clustering} developed a meta-learning-based framework that automatically selects the optimal clustering algorithm using a knowledge base, aligning closely with the principles underlying our AutoCluster methodology.

Addressing the topic modeling component, Boutaleb et al. \cite{boutaleb2024bertrend} introduced BERTrend, a neural topic modeling framework tailored to detect emerging trends in large-scale text corpora. BERTrend combines BERTopic embeddings with a temporal signal classification module to categorize trends as strong, weak, or noisy. This refined taxonomy enhances the tracking of topic trajectories and improves the accuracy of forecasting future relevance. Notably, the temporal signal classification component has been directly integrated into our system.

The authors of \cite{tarek2024} adopt a static machine learning approach—using a fixed set of models—to predict trending topics from scientific paper abstracts in response to user queries. In addition, they introduced a significantly improved pipeline, which they systematically applied to the software engineering domain as a comprehensive case study \cite{mahmoud2025predicting}. Their work served as the foundational basis for our proposed approach.

Despite these advancements, several gaps remain. The main limitations include the absence of clustering and dimensionality reduction, which leads to redundant or poorly defined topics, and the fact that many existing approaches do not fully automate topic modeling and labeling, or fail to integrate HPO into clustering and topic extraction. Furthermore, automation in time series analysis remains limited. These shortcomings highlight the need for a unified, end-to-end AutoML pipeline tailored for trend prediction tasks in a fully automated manner.

\section{Methodology}

\begin{figure*}[h!]
    \centering
    \includegraphics[width=\textwidth, height=0.2\textheight, keepaspectratio]{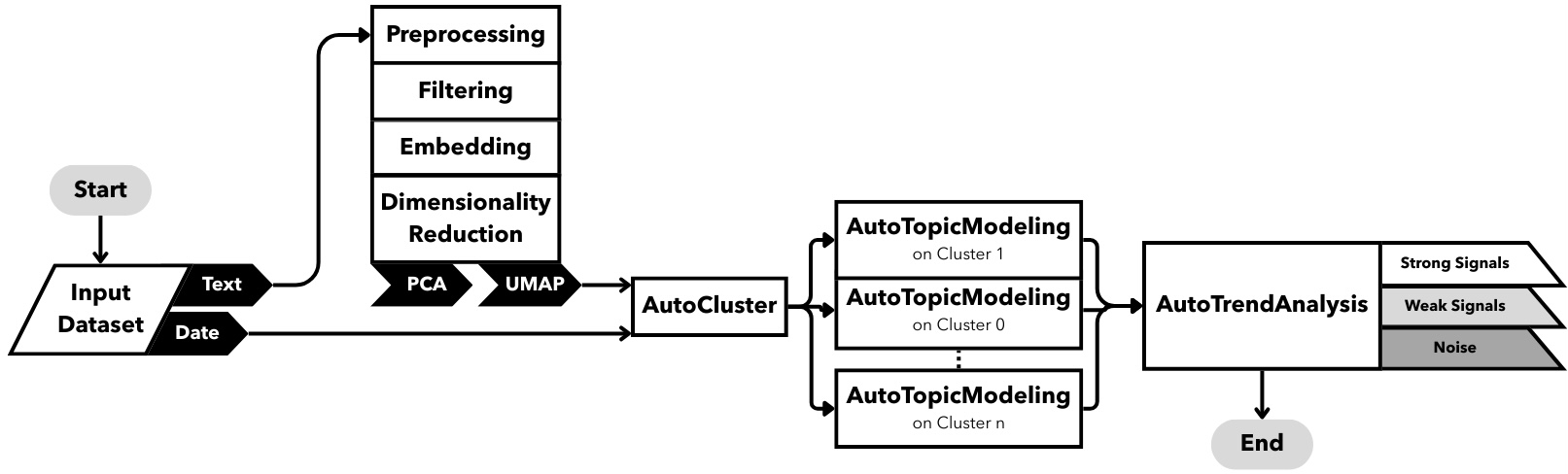}
    \caption{Methodology Diagram}
    \label{fig:methodology}
\end{figure*}

This section presents an overview of the
framework stages. Figure \ref{fig:methodology} shows the framework Methodology Diagram.

\subsection{Dataset Collection}

The dataset used in this study was compiled through a structured three-stage process to ensure relevance, quality, and scalability for trend forecasting in mobile network research.

\textbf{Stage 1: Identify Sources.} We began by querying the Scopus database using domain-relevant keywords such as “Wireless Networks”, “Wireless Communications”, “Mobile Computing”, and “Internet of Things". Filters were applied to restrict results to the subject area of \textit{Computer Networks \& Communications} and publication types limited to \textit{Journals} and \textit{Conference Proceedings}. This initial filtering yielded 33 distinct publication sources.

\textbf{Stage 2: Evaluate Sources.} To ensure high-quality input data, we retained only those sources with an H5-Index of 20 or higher. Furthermore, we restricted selection to well-established publishers, including IEEE, Elsevier, and ACM. This step was essential for maintaining academic rigor and credibility.

\textbf{Stage 3: Fetch Titles \& Abstracts.} We extracted metadata for papers published between 2010 and 2024, including titles and abstracts. Entries lacking abstracts were excluded to preserve semantic consistency during downstream processing. The final corpus comprised 127,820 records, providing a robust basis for topic modeling and trend analysis.

\subsection{Preprocessing and Filtering}

Before clustering, abstracts undergo several preprocessing steps to enhance semantic analysis. First, irrelevant named entities (such as persons, places, and institutions) are removed. Next, the text is normalized by converting it to lowercase, eliminating punctuation, short tokens, numbers, and stop words. Lemmatization is applied to consolidate word variations into their base forms. To minimize noise, excessively common tokens (appearing in \textgreater{}90\% of documents) and rare tokens (occurring fewer than twice) are filtered out. The final output is a refined, lemmatized set of tokens suitable for embedding and topic modeling.

\subsection{Embedding and Dimensionality Reduction}

To obtain a numerical representation of the semantic content of each abstract, we employed a document embedding model. Specifically, we utilized the \texttt{allenai-SPECTER} model, which is pre-trained on scientific literature, making it particularly well-suited for our domain-specific dataset. The model generates a 768-dimensional embedding vector for each abstract, ensuring that semantically similar abstracts are positioned closely in the resulting vector space.

Given the high dimensionality of the embeddings, using them directly in subsequent steps would be computationally inefficient. Therefore, we applied a two-stage dimensionality reduction process. Principal Component Analysis (PCA) was first used to denoise the embeddings and retain the components that capture the most variance. Subsequently, Uniform Manifold Approximation and Projection (UMAP) was applied to further reduce the dimensionality while preserving the global and local structure of the data. The resulting low-dimensional vectors were then appended to the original dataset and used in the downstream clustering tasks.

\subsection{AutoCluster}

Clustering was performed prior to topic modeling to ensure that the thematic structure within the dataset is preserved at a finer granularity. Without clustering, topic modeling on the entire corpus could lead to overly general topics that fail to capture domain-specific topics.

To obtain well-partitioned clusters for downstream topic modeling, we adopt a meta-learning approach that selects the most suitable clustering algorithm based on dataset meta-features. Specifically, we extract statistical meta-features such as mean, variance, skewness, kurtosis, mean correlation, and density variation.

\begin{figure}[h!]
    \centering
    \includegraphics[width=0.5\textwidth]{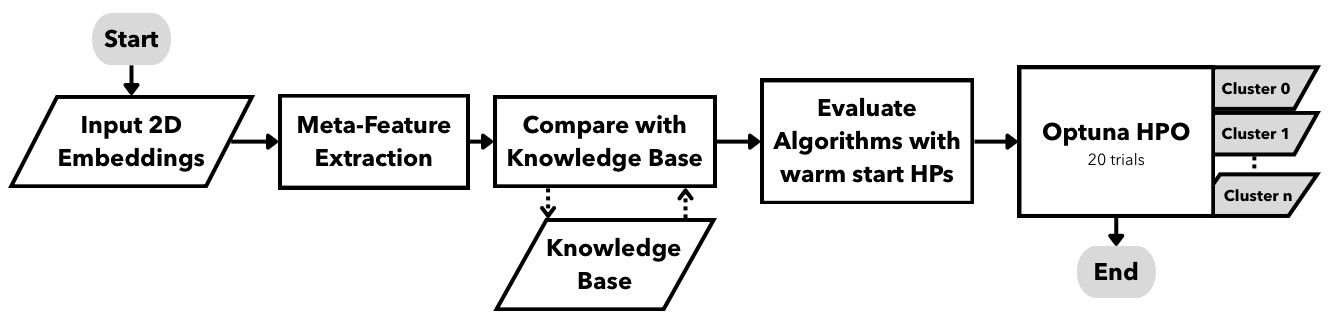}
    \caption{AutoCluster Diagram}
    \label{fig:autocluster}
\end{figure}

As illustrated in Figure \ref{fig:autocluster}, AutoCluster begins by extracting the six meta-features from the 2D embeddings and comparing them against a pre-populated knowledge base. This knowledge base was constructed using 1,000 synthetic datasets, each represented as a matrix of randomly generated numerical values, reflecting the numerical nature of embedding vectors. For each synthetic dataset, nine clustering algorithms were applied and optimized using Optuna, with the silhouette score serving as the objective function. Only valid results (Silhouette Score \textgreater{} 0.3, Davies-Bouldin index \textless{} 1.0, Number of Clusters between 3 and 15) were retained in the knowledge base. Each valid entry in the knowledge base stores the champion clustering algorithm, the corresponding six meta-feature values, and the optimized hyperparameters (HP).

For the Mobile Networks dataset, the top three entries from the knowledge base—determined using cosine similarity on the meta-feature vectors—are retrieved. The corresponding clustering algorithms, along with their associated hyperparameters, are then evaluated on the Mobile Networks dataset as warm-start configurations. The best-performing algorithm among these candidates proceeds to the optimization phase, where we conduct 20 Optuna trials to fine-tune its hyperparameters. The silhouette score is used as the objective function to be maximized during this optimization process. This approach enables a generalized clustering procedure that remains robust to variations in the structure or distribution of the input Mobile Networks dataset, ensuring adaptability to evolving data characteristics.

\subsection{AutoTopicModeling}

Applying high-quality topic modeling to large datasets is computationally intensive, particularly when evaluating multiple models that may perform differently across various data segments. To address this, we employ a successive halving approach, which significantly reduces computational overhead by iteratively eliminating the least promising models based on their coherence scores. This strategy allows us to focus resources on the most effective models while maintaining high-quality topic extraction.

\begin{figure*}[h!]
    \centering
    \includegraphics[height=0.13\textheight]{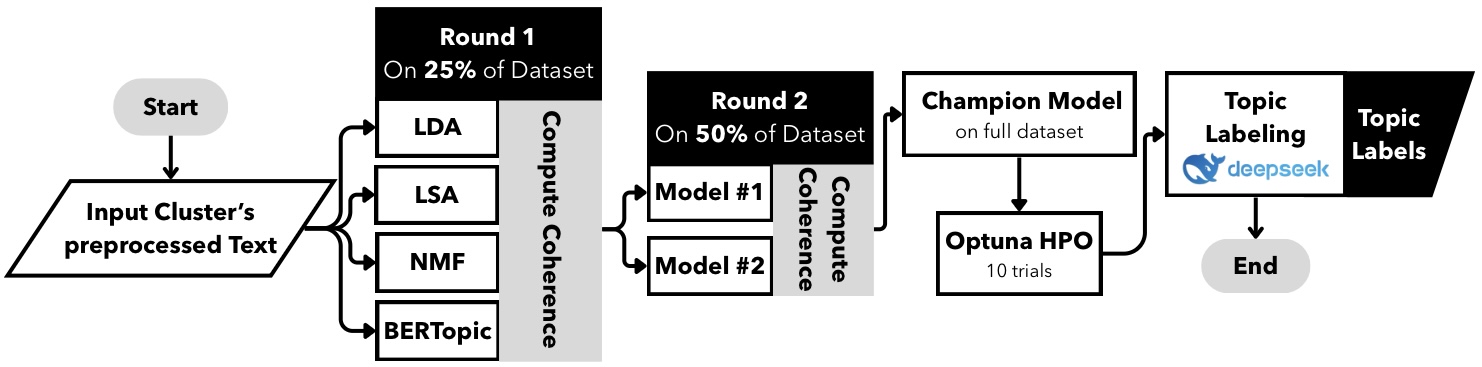}
    \caption{AutoTopicModeling Diagram}
    \label{fig:tm}
\end{figure*}

As illustrated in Figure \ref{fig:tm}, the topic modeling process begins by evaluating four candidate models—Latent Dirichlet Allocation (LDA), Nonnegative Matrix Factorization (NMF), Latent Semantic Analysis (LSA), and BERTopic—on a 25\% subset of each cluster. Based on coherence scores, the two best-performing models advance to a second round using 50\% of the data. The champion model from this round is then selected for HPO using Optuna, with 10 trials aimed at maximizing the coherence score. The final output consists of the top 12 representative topic words per cluster, which are appended to the original dataset in a new \texttt{topic} column.

To obtain meaningful labels, we employ the DeepSeek API to each list of topic words, using the following prompt: \texttt{"Analyze these topic words generated from search subject: '{subject}'.
For each topic, provide ONE concise label (2 to 6 words)"}, where \texttt{'{subject}'} is the subject of the dataset.

If the number of clusters is excessively high, an optional post-labeling step can be applied to reduce the number of distinct topic labels, depending on the specific use case. To achieve this, we leverage the DeepSeek API once again to generalize semantically similar labels. The topic labels generated in the previous step are submitted to the API with a structured prompt designed to group highly similar topics under broader, domain-relevant categories. The prompt is formulated as follows:

\texttt{"Generate mappings for these topics by merging extremely similar topic labels into more general labels in the domain of \texttt{'{subject}'}. The total number of distinct labels should be between \texttt{'x'} and \texttt{'y'}.}

This step ensures more concise and interpretable labeling, which is especially useful for downstream analysis or visualization.

This stage incorporates Large Language Model (LLM)-based capabilities including: (1) Semantic Merging of near-duplicate topics; (2) Contextual Understanding informed by the subject domain; (3) Domain-Aware Relatedness for disambiguating polysemous terms.

Final labels are appended to the original dataset in a new column \texttt{topic\_label}. This end-to-end pipeline enables scalable, efficient, and context-aware topic modeling for each cluster, effectively balancing computational robustness with semantic interpretability.

\subsection{AutoTrendAnalysis}

\begin{figure*}[h!]
    \centering
    \includegraphics[height=0.12\textheight]{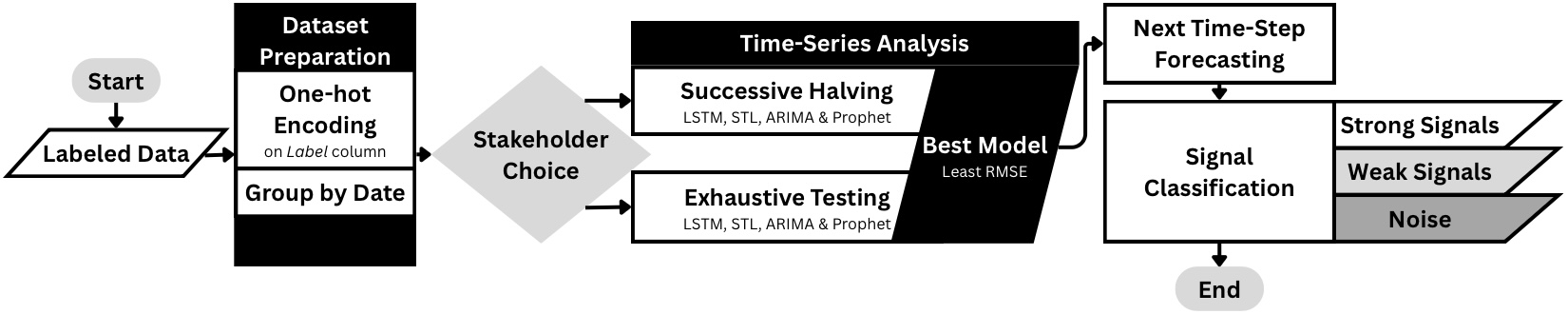}
    \caption{AutoTrendAnalysis Diagram}
    \label{fig:ta}
\end{figure*}

The final stage of the pipeline, \textbf{AutoTrendAnalysis}, incorporates temporal dynamics into topic evolution by forecasting future trends using either classical or neural time-series models. Building on the previously clustered and labeled data, this component predicts the popularity trajectory of each topic over time, offering valuable insights into emerging and declining themes (Figure~\ref{fig:ta}).

\subsubsection{Data Preparation}

Topic labels are transformed into one-hot encoded binary indicator columns, where each row marks the presence of specific topics within a document. These indicators are then aggregated by \texttt{date}, summing the counts to generate time series of topic frequencies. The final dataset includes only the \texttt{date} and corresponding topic columns, organized chronologically and indexed by date. This structure is compatible with various time series forecasting models, including ARIMA, STL, Prophet, and LSTM, which we use.

\subsubsection{Model Selection Strategies}
Two distinct evaluation strategies are supported for model selection:

\begin{itemize}
    \item \textbf{Exhaustive Evaluation (default)}: Each of the four candidate forecasting models is trained on the initial 80\% of the time series and evaluated on the remaining 20\%. Model performance is quantified using the RMSE.

    \item \textbf{Successive Halving}: The methodology employed in \textbf{AutoTopicModeling} is reutilized in this stage to ensure efficiency in model evaluation. The top-performing model, based on RMSE, is then retrained on the full dataset with an 80-20 train-test split for final evaluation.
\end{itemize}

\subsubsection{Next Time-Step Forecasting}

The champion model generates forecasts for the number of scientific papers associated with each topic in the upcoming time step, with the temporal granularity—weekly, monthly, or yearly—configurable to suit specific application needs. This flexibility allows the forecasting process to adapt to varying temporal resolutions depending on the use case.

\subsubsection{Signal Classification}

Forecasted topic values are classified into:

\begin{itemize}
    \item \textbf{Strong Signals}: Values above the 40th percentile, signaling potential long-term relevance or growing importance.
    \item \textbf{Weak Signals}: Values between the 10th and 40th percentile with positive slopes, indicating potential future growth.
    \item \textbf{Noise}: Values below the 10th percentile or non-increasing trends, indicating diminishing long-term importance
\end{itemize}

This categorization emphasizes topics based on their trajectory and significance, enabling data-driven insights for strategic decision-making. By integrating robust statistical methods with meaningful semantic analysis, the pipeline generates trend forecasts that are both interpretable and practically applicable for downstream tasks.

\section{Experimentation and Results}
\subsection{Experiment 1: AutoCluster}

In this experiment we apply the \textbf{AutoCluster} procedure on the "Wireless Networks and Mobile Computing" dataset to effectively cluster the dataset.

The top three most similar entries retrieved from the knowledge base included two instances of Gaussian Mixture Model (GMM) clustering and one of Agglomerative Clustering. Among these, an instance of GMM achieved the highest silhouette score (which measures how well-separated and cohesive the clusters are) during the inference phase, with a value of 0.347, qualifying it for the optimization stage.

Following HPO with Optuna, the GMM yielded an improved silhouette score of 0.3860 and a Davies-Bouldin index (which evaluates cluster compactness and separation, where lower values are better) of 0.777. The optimal configuration identified was \texttt{n\_components = 15} and \texttt{covariance\_type = 'tied'}.

These results indicate that GMM was well-suited for capturing the latent structure of the \textit{Mobile Networks} dataset under the given embedding space. Additionally, the Davies-Bouldin index of 0.777 further supports the quality of clustering, indicating relatively well-defined and distinct clusters. These metrics demonstrate that the AutoCluster framework effectively identifies and tunes appropriate models based on data-specific characteristics.

\subsection{Experiment 2: AutoTopicModeling}

In this experiment, \textbf{AutoTopicModeling} was applied to each cluster to determine and assign topic labels to individual abstracts within the \textit{Mobile Networks} dataset.

Across all clusters, the minimum coherence score observed was 0.442, indicating a strong semantic alignment between the abstracts and their corresponding topic keywords. The champion model varied across clusters, with LDA and NMF dominating most cases. Notably, a single cluster selected BERTopic as its top-performing model, highlighting the framework’s flexibility in adapting to local topic structures.

Owing to the large number of clusters, the initial topic modeling phase produced 170 distinct labels across the dataset. Several of these labels were highly specific—for example, \textit{Millimeter Wave Beamforming} and \textit{RFID Anti-Collision Protocol}—which, while precise, limited broader interpretability.

To enhance usability and generalization, the optional label merging step was applied using the DeepSeek API. This process consolidated semantically similar labels, resulting in a reduced and more manageable set of 49 distinct labels, better suited for downstream analysis and application. Resulting merged topic labels are displayed in Table \ref{tab:trends}.

\subsection{Experiment 3: AutoTrendAnalysis}

In this experiment, the \textbf{AutoTrendAnalysis} module was employed to categorize topics into three signal classes: \textit{strong}, \textit{weak}, and \textit{noise}.

Given that computational efficiency was not a primary constraint, we selected the exhaustive evaluation strategy over successive halving. All forecasting models were trained and assessed on the complete dataset. Among them, STL emerged as the top-performing model, achieving the lowest RMSE of 36.76, indicating high prediction accuracy with minimal error.

The resulting topic classifications are presented in Table~\ref{tab:trends}.

\subsection{Discussion}

The forecasted trends offer a concise yet insightful snapshot of current and emerging research priorities in Mobile Networks as of January 2025. Strong signals such as \textit{AI \& Learning Algorithms}, \textit{Edge Computing}, and \textit{Smart Devices \& IoT} emphasize the field's shift toward intelligent, decentralized, and adaptive systems—key characteristics of future 6G networks.

Core mobile networking topics like \textit{Channel Modeling \& Estimation}, \textit{Routing Protocols}, and \textit{MIMO \& Beamforming} remain central, reflecting sustained interest in optimizing wireless communication efficiency and reliability. Meanwhile, the prominence of \textit{Network Security}, \textit{Data Privacy}, and \textit{Secure Communication} highlights ongoing concerns over protecting distributed infrastructures and devices.

Several interdisciplinary applications, such as \textit{Computer Vision}, \textit{Healthcare}, and \textit{Finance}, although not exclusive to mobile networking, appear strongly due to their reliance on mobile connectivity and edge intelligence for real-time processing and delivery.

Weak signals, including \textit{Big Data} and \textit{Algorithm Design}, likely represent backend enablers rather than front-facing innovations. Some noise signals—such as \textit{Caching}, \textit{Multiple Access Techniques}, and \textit{Spectrum Management}—while traditionally relevant, may be experiencing a research plateau or integration into broader system-level studies.

The RMSE of 36.76 supports the reliability of these predictions. Overall, the trend taxonomy provides academics with a practical map of impactful directions in Wireless Networks and Mobile Computing research, particularly at the intersection of AI, secure infrastructure, and emerging applications.

\section{Conclusion, Limitations and Future Work}

This work introduces a scalable pipeline for forecasting research trends in mobile networks by integrating semantic embeddings, meta-learning-driven clustering, adaptive topic modeling, and time-series forecasting. Using over 127,000 abstracts, we employed SPECTER for embedding, AutoCluster for algorithm selection, and successive halving for efficient topic and model evaluation.

\begin{table}[H]
\caption{Forecasted Trends in Mobile Networks}
\centering
\renewcommand{\arraystretch}{1.0}
\begin{tabular}{|c|c|}
\hline
\multicolumn{2}{|c|}{\textbf{Date: 01-01-2025}} \\
\hline
\textbf{Topic} & \textbf{Forecasted Docs} \\
\hline
\multicolumn{2}{|c|}{\textbf{Strong Signals}} \\
\hline
AI \& Learning Algorithms & 1233.76 \\
Network Security & 823.51 \\
Edge Computing & 762.69 \\
Energy \& Power Management & 660.32 \\
Computer Vision & 462.62 \\
Channel Modeling \& Estimation & 382.36 \\
Vehicular Networks & 365.23 \\
Smart Devices \& IoT & 342.65 \\
Sensing \& Recognition Systems & 307.92 \\
Data Privacy \& Encryption & 305.01 \\
Healthcare \& Biomedical Applications & 302.86 \\
Sensor Networks & 301.12 \\
Natural Language Processing & 286.81 \\
UAV \& Aerial Networks & 281.60 \\
Routing \& Transmission Protocols & 271.66 \\
Applications in Finance & 266.69 \\
Secure Communication & 262.94 \\
Signal Processing & 254.35 \\
Resource Management & 253.99 \\
Wireless Networks & 250.19 \\
Network Architecture \& Virtualization & 234.95 \\
Traffic Analysis \& Prediction & 227.91 \\
Reflective \& Reconfigurable Surfaces & 208.49 \\
Blockchain \& Distributed Systems & 202.33 \\
Applications in Education & 178.65 \\
Network Performance \& Optimization & 177.88 \\
Localization \& Positioning & 171.23 \\
Mobile Networks & 162.02 \\
MIMO \& Beamforming & 158.19 \\
\hline
\multicolumn{2}{|c|}{\textbf{Weak Signals}} \\
\hline
Algorithm Design \& Optimization & 140.91 \\
Big Data \& Cloud Computing & 124.41 \\
Optical Communication & 99.88 \\
Software \& System Security & 90.12 \\
General Research Methods & 81.93 \\
\hline
\multicolumn{2}{|c|}{\textbf{Noise Signals}} \\
\hline
Control \& Adaptation Algorithms & 154.14 \\
Multiple Access Techniques & 140.12 \\
Spectrum Management & 109.77 \\
Caching \& Data Delivery & 108.56 \\
Antenna Design \& Hardware & 85.35 \\
Satellite \& Hybrid Networks & 84.36 \\
Fault Detection \& Reliability & 77.91 \\
Interference \& Spectrum Management & 69.81 \\
Crowdsensing \& Participation & 68.62 \\
Underwater Communication & 68.08 \\
Error Correction \& Coding & 64.03 \\
Robotics \& Control & 51.78 \\
Data Collection \& Sensing & 40.44 \\
Multimedia \& Streaming & 27.31 \\
QoS \& User Experience & 10.92 \\
\hline
\multicolumn{2}{|c|}{\textbf{RMSE:} 36.76} \\
\hline
\end{tabular}
\label{tab:trends}
\end{table}

The pipeline effectively surfaced dominant and emerging themes such as AI optimization, edge computing, and security, achieving a forecasting RMSE of 36.76. These results confirm the pipeline's robustness and its value for understanding evolving research landscapes in mobile networking.

\subsection{Limitations}

Despite promising results, the proposed pipeline has a few limitations. First, it relies on static SPECTER embeddings, which do not capture shifts in language or meaning over time. Second, topic labeling via large language models, while effective, may introduce inconsistencies or ambiguity. Lastly, the forecasting component focuses on short-term predictions and does not model long-range trends or quantify uncertainty.

\subsection{Future Work}

Key areas for future improvement include incorporating temporal embeddings to capture semantic drift and deploying the system for real-time trend monitoring. Additionally, integrating interpretability tools can enhance forecast transparency, while applying the pipeline across different domains and languages will help validate its generalizability.

\bibliographystyle{IEEEtran}
\bibliography{references}
\end{document}